\documentclass[twocolumn]{revtex4}
\usepackage[T1]{fontenc}
\usepackage{epsfig}
\usepackage{amsmath}
\usepackage{amssymb}
\usepackage{color}
\usepackage{epstopdf}
\usepackage{ifpdf}

\renewcommand{\Re}{\mathrm{Re}\,}

\renewcommand{\i}{\mathrm{i}}
\newcommand{\e}{\mathrm{e}}

\newcommand{\m}{\mathcal}
\newcommand{\ph}{\varphi}

\newcommand{\vid}[1]{\langle #1\rangle}

\begin{document}
\title{Negative flow of energy in a mechanical wave}
\author{A.~Matulis}\email{matulisalg@gmail.com}
\affiliation{Semiconductor Physics Institute, Center of Physical
Sciences and Technology, Saul\.{e}tekio 3, LT-10222 Vilnius, Lithuania}
\author{A.\ Acus}\email{arturas.acus@tfai.vu.lt}
\affiliation{Institute of Theoretical Physics and Astronomy, Vilnius University,
Saul\.{e}tekio 3, LT-10222 Vilnius, Lithuania}

\begin{abstract}
A classical system, which is analogous to the quantum one with a
backflow of probability, is proposed. The system consists of a
chain of masses interconnected by springs, as well attached by
other springs to fixed supports. Thanks to the last springs the
cutoff frequency and dispersion appears in the spectrum of waves
propagating along the chain. It is shown that this dispersion
contributes to the appearance of a backflow of energy. In the case
of the interference of the two waves, the magnitude of this
backflow is an order of magnitude higher than the value of the
probability backflow in the mentioned quantum problem. The
equation of Green's function is considered, and it is shown that
the backflow of energy is also possible when the system is excited
by two consecutive short pulses. This classical backflow
phenomenon is explained by the branching of energy flow to local
modes, what is confirmed by the results for the forced damped
oscillator. It is shown that even in such a simple system the
backflow of energy takes place (both an instantaneous and on
average).

\vspace{1.5mm}\noindent \textbf{Keywords:} Wave, dispersion, energy, backflow, 1D lattice, oscillator
\end{abstract}

\maketitle

\section{Introduction}
There has always been an interest in essential quantum phenomena without classical analogues. The negative probability flow was believed to be such case, which was first mentioned by Allcock~\cite{Al69} when he considered the time of arrival in quantum mechanics. Later this phenomenon was described in detail by Bracken and Melloy~\cite{Bra94}. Exhaustive survey of quantum models and a broad review of the literature pertaining to this problem can be found in Yearsley's \textit{et al} paper~\cite{Ye12} and Goussev's papers~\cite{goussev2020,goussev2008}. The problem of negative flow was also considered in the field of optical phenomena~\cite{Berry2010}. There have been attempts to show that the effect of the negative flow may take place when replacing the quantum description with the classical one (see, for instance, Villanueva's paper~\cite{Vill20}).

Goussev's paper~\cite{goussev20} asserts that the effect of the negative probability flow occurs not only for a free quantum particle, but also for an ensemble of free classical particles with a Gaussian distribution of positions and momenta. This is quite expectable because the particle ensemble always shows a greater variety of motion types than the movement of an individual particle~\cite{mat19}.

We believe that the phenomenon of the negative energy flow is inherent to
any wave equation, whether it is related to the quantum model or
to the classic one. The only important condition is the presence of dispersion
what is most often due to local degrees of freedom in the media where
the waves propagate. The aim of this paper is to demonstrate that the negative
energy flow has to take place in a simple mechanical system: the chain of masses
interconnected by harmonic springs, and its continuous version -- the
waves in elastically braced string. In addition, we consider the
finite version of the above chain in the limit case of short length converting
itself into the problem of the harmonic oscillator.

The paper is organized as follows. In Section 2 a description of the basic model is provided,
and in Section 3 the energy flows are introduced and the continuous limit of the model
is formulated. In Section 4 the energy flow of the individual sinusoidal wave is calculated,
and in Section 5 it is shown that in the case of two sinusoidal waves
the local negative energy flow becomes possible. In Section 6 we consider the Green's
function behavior and show that the string demonstrates the local negative energy flow
when it is excited by two consecutive pulses. In section 7 we deal with a harmonic
oscillator as the ultimate case of a super-short chain, which actually demonstrates
the analog to the backflow of energy. Our conclusions are presented in Section 8.

\section{Model}

\begin{figure}[!ht]
  \includegraphics[width=70mm]{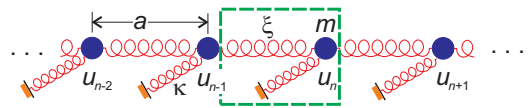}
\caption{Basic 1D lattice. Its primitive cell is shown by green dashed rectangular.}
\label{fig1}
\end{figure}

We consider simple 1D (one dimensional) mechanical system shown in Fig.~\ref{fig1}.
It consists of point masses $m$ (indicated by filled circles) spaced a distance $a$
apart and interconnected by linear springs of stiffness $\xi$. Each mass is also
attached to other spring of stiffness $\kappa$. The ends of these additional
springs are kept at fixed positions. These extra springs provide us with the necessary
wave dispersion.
The masses are allowed to move in the horizontal
direction only, and the symbols $u_n$ stand for their coordinates: the deviations
from the equilibrium positions.

The dynamics of this 1D lattice is characterized by the following Hamiltonian:
\begin{equation}\label{ChHam}
  H = \frac{1}{2}\sum_{n=-\infty}^{\infty}\left[p_n^2/m + \kappa u_n^2
  + \xi\left(u_n-u_{n-1}\right)^2\right],
\end{equation}
where the symbol $p_n$ stands for the canonical momentum, conjugate to each coordinate $u_n$.
The resulting equations of motion are
\begin{subequations}\label{ChJud}
\begin{eqnarray}
\label{ChJud1}
  \dot{u}_n &=& = p_n/m, \\
\label{ChJud2}
  \dot{p}_n &=& m\ddot{u}_n = - \kappa u_n + \xi\left(u_{n-1}
  - 2u_n + u_{n+1}\right).\phantom{m}
\end{eqnarray}
\end{subequations}

\section{Energy and its flow}
\label{sec_ensr}

Energy and its flow appear naturally, when calculating the first integral
of differential motion equations. Indeed, multiplying Eq.~(\ref{ChJud2})
by $\dot{u}_n$ we rewrite it as follows:
\begin{equation}\label{ruten}
  \frac{d}{dt}\left(\frac{m}{2}\dot{u}_n^2 + \frac{\kappa}{2}u_n^2\right)
  = \xi\dot{u}_n\left(u_{n-1}-2u_n+u_{n+1}\right).
\end{equation}
The expression in parenthesis in the left hand side of this equation looks like energy of
the primitive cell. Consequently, the right hand side of it has to be related to the energy flow.
In order to make a final decision, however, it is necessary to take into account the
potential energy of the spring which connects neighboring masses $\xi(u_n-u_{n-1})^2/2$
(see, the Hamiltonian (\ref{ChHam})).
The above equation then turns into the following one:
\begin{equation}\label{StSraut}
\begin{split}
    \dot{\m{E}}_n =& \frac{d}{dt}\left[\frac{m}{2}\dot{u}_n^2 + \frac{\kappa}{2}u_n^2
  + \frac{\xi}{2}\left(u_n-u_{n-1}\right)^2\right] \\
   =& \xi\left[\dot{u}_{n-1}\left(u_{n-1}-u_n\right) - \dot{u}_n\left(u_n-u_{n+1}\right)\right].
\end{split}
\end{equation}
Actually it is the energy conservation rule applied to the primitive cell shown in
Fig.~\ref{fig1} by dashed rectangle \footnote{It is worth noting that the choice of
a primitive cell is rather arbitrary, but it does not affect our further conclusions.}.
It is obvious that the energy of the chosen primitive cell changes because a certain
amount of energy flows into it, or flows out. These energy flows are determined
by the expression of the right hand side of Eq.~(\ref{StSraut}).
In this way, we will determine the flow of energy along the chain as the energy flowing
through the right edge of the mentioned primitive cell (or through mass at $n$):
\begin{equation}\label{dsraut}
  J_n = \xi\dot{u}_n\left(u_n-u_{n+1}\right),
\end{equation}
that actually means the work done by mass $m$ in a unit of time over the spring attached
to this mass on its right side. In this case, the value of $J_{n-1}$
can be seen as the flow of energy
entering the primitive cell on its left side, and Eq.~(\ref{StSraut}) itself
determines the energy balance of this primitive cell.

Due to the linearity of the equation (\ref{ChJud2}), its solution can be presented as an
exponential running wave
\begin{equation}\label{ekspo}
  u_n = \e^{\i(kan - \omega t)},
\end{equation}
satisfying the following dispersion relation:
\begin{equation}\label{spec}
  \omega = \sqrt{4\omega_0^2\sin^2\left(ka/2\right) + \Omega^2},
\end{equation}
where
\begin{equation}\label{konstdef}
  \Omega^2 = \kappa/m \qquad \text{and} \qquad \omega_0^2 = \xi/m.
\end{equation}
In Eq.~(\ref{spec}) we see two causes of dispersion. Namely, the first constituent under the
radical appears because of the discrete nature of the model, while the second one is due
to the presence of a cutoff frequency $\Omega$, leading to a forbidden band in the frequency
spectrum.

We will focus mainly on the second type of dispersion that comes from the cutoff frequency
and allows us to introduce a simpler continuous version of the problem,
making the transition to the limit $a\to 0$.
Thus, replacing the mass number by the coordinate $x = na$, and the energy of the primitive
cell by the density of energy $\m{E}(x) = \m{E}_n/a$, introducing new constants
\begin{equation}\label{ribzym}
   \rho = m/a, \quad G = \kappa/a, \quad T = \xi a, \quad c = a\omega_0 = \sqrt{T/\rho},
\end{equation}
and variables
\begin{equation}\label{mast}
\begin{split}
&  t\to\Omega^{-1}t, \quad x\to c\Omega^{-1}x, \\
&  \m{E}\to\rho\,\Omega^2\m{E}, \quad J\to c\rho\,\Omega^2J,
\end{split}
\end{equation}
we replace Eqs.~(\ref{ChJud2}), (\ref{StSraut}) and (\ref{dsraut}) by the
following more simple and convenient system of equations:
\begin{subequations}\label{bdlg}
\begin{eqnarray}
\label{bdlg1}
  u_{tt} &=& u_{xx} - u, \\
\label{bdlg2}
  \m{E}(x) &=& \frac{1}{2}\left(u_t^2 + u^2 + u_x^2\right), \\
\label{bdlg3}
  J(x) &=& -u_tu_x.
\end{eqnarray}
\end{subequations}
In fact, the dispersion relation of Eq.~(\ref{bdlg1})
\begin{equation}\label{lwspec}
  \omega = \sqrt{k^2 + 1}
\end{equation}
is the long wave approximation ($ka \ll 1$) of Eq.~(\ref{spec}).
The set (\ref{bdlg}) of dimensionless equations will be our main tool.
Eq.~(\ref{bdlg1}) is well known one-dimensional Klein-Gordon equation, and it also describes the
waves in elastically braced string \cite{mors53}. It was discussed
in detail in A.~Matulis paper \cite{ma99}.

\section{Energy flow in an individual wave}
\label{sec_indbang}

First we consider the energy flow in the case of an individual wave, running to the right
along the $x$-axis. We present its amplitude in the following real form:
\begin{equation}\label{ekstol}
  u(x,t) = \cos\ph, \qquad \ph = kx-\omega t,
\end{equation}
because the energy and its flow that interest us are the square functions of the wave
amplitude, and thus, the complex exponent form isn't convenient.
The amplitude (\ref{ekstol}) satisfies Eq.~(\ref{bdlg1}) automatically if the frequency $\omega$ and
the wave number $k$ are related by Eq.~(\ref{lwspec}).

Substituting amplitude (\ref{ekstol}) into Eqs.~(\ref{bdlg2},c), we find expressions
of energy
\begin{equation}\label{bvt}
\begin{split}
  \m{E} &= \frac{1}{2}\left[(\omega^2 + k^2)\sin^2\ph + \cos^2\ph\right] \\
    &= \frac{\omega^2}{2} - \frac{k^2}{2}\cos(2\ph),
\end{split}
\end{equation}
and its flow
\begin{equation}\label{bvsr}
  J = \omega k\sin^2\ph.
\end{equation}
When interested in mean values, the averages of energy density and flow over the period
of rapidly changing phase $\ph$:
\begin{equation}\label{aver}
  \vid{E} = \omega^2/2, \qquad \vid{J} = \frac{k\omega^2}{2\omega} = v_{\mathrm{gr}}\vid{E}
\end{equation}
are used. Here the symbol
\begin{equation}\label{grvel}
  v_{\mathrm{gr}} = d\omega/dk = k/\omega
\end{equation}
stands for the group velocity.

We see that although mean values show a fairly monotonous picture (the uniform
wave energy is transferred with a group velocity), the individual wave demonstrates
the oscillating flow of energy,
and there are points (corresponding to phase values $\ph_n=\pi n$) where this flow
is zero. Therefore, it is hoped that the superposition of several such waves
could show even a negative local flow. In the next section, we'll consider the
two-wave superposition, which, in the case of Schrodinger's 1D equation,
demonstrated the negative flow of probability \cite{Bra94}.

\section{Superposition of two waves}

Let us consider the following superposition of two waves
\begin{equation}\label{dvbgrl}
  u(x,t) = \cos\ph_1 + A\cos\ph_2,
\end{equation}
where the symbol
\begin{equation}\label{twophase}
  \ph_i \equiv \ph_i(x,t) = k_ix-\omega_it
\end{equation}
stands for the individual wave phase.
According to the equation (\ref{bdlg3}) it corresponds to the following flow of energy
(for the sake of convenience divided by $4\omega_1k_1/\pi^2$):
\begin{equation}\label{ensrdvir}
  J_0 = \frac{\pi^2 J}{4\omega_1k_1} = \frac{\pi^2}{4}\left(\sin\ph_1 + f\sin\ph_2\right)
  \left(\sin\ph_1 + g\sin\ph_2\right),
\end{equation}
where
\begin{equation}\label{gdzdif}
  f = \frac{\omega_2}{\omega_1}A, \qquad
  g = \frac{k_2}{k_1}A = \frac{v_{\mathrm{gr},2}}{v_{\mathrm{gr},1}}f.
\end{equation}

We see that the dispersion is really important, because in the case of its absence the group
velocities are equal ($v_{\mathrm{gr},1}=v_{\mathrm{gr},2}$). Consequently,
$f=g$, the expression on the right hand side of
Eq.~(\ref{ensrdvir}) becomes a function squared, and the energy flow can not be negative.

Our goal is to identify the $f$ and $g$ values for which there is a
backflow in the widest interval for phases and the definition of
the limits of this flow itself. Let's keep in mind that because of
the periodicity of flow (\ref{ensrdvir}) it is enough to consider
it in the finite square area $0 \leqslant \ph_1,\,\ph_2 < 2\pi$.
In addition, the backflow is possible only in the case when the
parentheses in Eq.~(\ref{ensrdvir}) are of different signs.
Obviously, this can only happen when trigonometric functions
$\sin\ph_1$ and $\sin\ph_2$ also have different signs. This
reduces the search area to two smaller squares. We choose one of
them $0 < \ph_1 < \pi$, $\pi < \ph_2 < 2\pi$, because the another
turns out to be a simple replacement of variables $\ph_i\to 2\pi -
\ph_i$ and leads to a similar result. For convenience, we will
change one of the phases: $\ph_2\to 2\pi - \ph_2$. Then the flow
which we are interested in is defined in $0 \leqslant \ph_1,\ph_2
< \pi$ square, and reads
\begin{equation}\label{baz0}
  J_0 = \frac{\pi^2}{4}\left(\sin\ph_1 - f\sin\ph_2\right)\left(\sin\ph_1 - g\sin\ph_2\right).
\end{equation}
This expression is symmetric with respect to the replacement of variables: $\ph_i \to \pi - \ph_i$.
Therefore, the search area can be narrowed down to the next square:
\begin{equation}\label{kvadratas}
  0 \leqslant \ph_1,\ph_2 < \pi/2.
\end{equation}

Note that the sinus functions in the above interval are the monotonously growing ones.
So one can replace them by more simple variables as follows:
\begin{equation}\label{fazikint}
  \sin\ph_1 = 2\xi/\pi, \quad \sin\ph_2 = 2\eta/\pi.
\end{equation}
This allows us to replace the flow of energy (\ref{baz0}) by the following polynomial:
\begin{equation}\label{para}
  J_0 = \left(\xi - f\eta\right)\left(\xi - g\eta\right) = F\cdot G
\end{equation}
which we have to consider in $0 \leqslant \xi,\eta\leqslant 1$ square.

\begin{figure}[!ht]
\includegraphics[width=50mm]{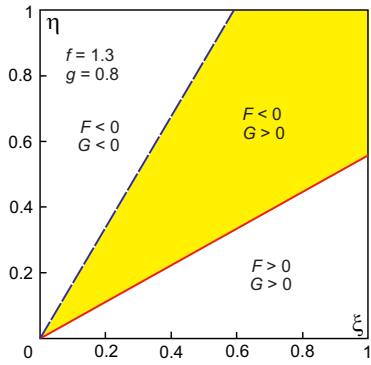}
\caption{The area of the negative energy flow.}
\label{fig2}
\end{figure}

Now it is very easy to set the boundaries of the negative flow area: one just need to find
the lines where this flow is zero. Since the expression (\ref{para}) consists of a product
of two factors, it is zeroed when any of them is zero. In this way, we define two
oblique straights, shown in Fig.~\ref{fig2} by red solid and blue dashed lines.
Each of them separates the area where the corresponding factor ($F$ or $G$) is positive,
from the area where it is negative. These two straights together with the edges of the square
defines the area shown in yellow, where the backflow exists.
From equation
\begin{equation}\label{lglin}
  F = \xi - f\eta = 0
\end{equation}
and Fig.~\ref{fig2}, it is obvious that the
smaller the coefficient $f$, the higher the straight.
So to have the maximum backflow area, we need to make as much difference as possible
between values of $f$ and $g$ coefficients. For instance, we can choose
\begin{equation}\label{fginf}
  f \to \infty, \quad g = 0.
\end{equation}
In this case the flow becomes $J_0 \sim -f\xi\eta$, and it is negative on the whole square
of interest.
However, according to the definitions (\ref{gdzdif})
in order to realize this favorable situation, we need to meet the implementation of the
following inequalities:
\begin{equation}\label{metasal}
  \omega_2 > \omega_1, \quad k_2 < k_1,
\end{equation}
what is actually the definition of meta-material \cite{ves64}.
This points to the close relationship of the local backflow of energy under consideration
with the popular area of the design and research of meta-materials, for which this
backflow is the main property.

\begin{figure}[!ht]
\includegraphics[width=50mm]{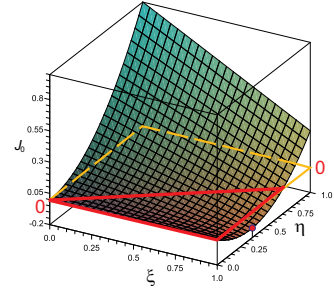}
\caption{Energy flow in the case of dispersion relation (\ref{lwspec}).}
\label{fig3}
\end{figure}

Restricting our consideration with the conventional models, we will take into account
the dispersion relation given by Eq.~(\ref{lwspec}). In this case, the ability to have
a local backflow can be evaluated as follows. Since the dispersion for the backflow problem
is essential, it makes sense to choose the wave vectors from the interval $\{0,1\}$
where this dispersion is the largest. So taking into account Eq.~(\ref{lwspec}) let's choose:
\begin{equation}\label{prpas}
  k_1 \to 0, \quad \omega_1 = 1, \quad k_2 = 1, \quad \omega_2 = \sqrt{2}.
\end{equation}
Then the parameters and flow normalized by the largest positive value will be as follows:
\begin{equation}\label{kprrez}
  f = \omega_2/\omega_1 = \sqrt{2}, \quad g \to \infty, \quad
  J_0 = \eta\left(\eta - \xi/\sqrt{2}\right).
\end{equation}
This flow is shown in Fig.~\ref{fig3}.
The zero plane is highlighted by a yellow rectangle, and the red straights
indicate the sector of backflow.
We see that the maximum backflow value is reached on the right side of the square
(indicated by red circle) at the point $\xi_0=1$, $\eta_0=1/2\sqrt{2}$. This value
reads
\begin{equation}\label{flmin}
  J_{0,\mathrm{max}} = - 0.125,
\end{equation}
what is about an order of magnitude more than similar values received for a free
electron in a quantum description \cite{Bra94}.

Considering wave processes scientists are often interested not only in the instant values of
wave characteristics, but also in their means: the values averaged by period.
In the case of the superposition of the two waves, it makes sense to consider the waves of
close frequencies. In this case it can be expected that these characteristics, averaged over
a fast process, still contain useful information about slow beats in both temporal and
coordinate dependencies. To illustrate this possibility, let's look at the superposition
of two waves with close to one frequencies ($\omega_1,\omega_2\sim 1$), and
\begin{equation}\label{k12parinkt}
  k_1 = 0, \quad k_2 \equiv k \ll 1,
\end{equation}
and the amplitude of this wave reads
\begin{equation}\label{twmamp}
  u(x,t) = A\cos t + \cos(kx - t).
\end{equation}
The energy flow of this wave is
\begin{equation}\label{twmflow}
  J = - u_xu_t = k\sin(kx-t)\left[-A\sin t + \sin(kx-t)\right].
\end{equation}
Now, with the average trigonometry functions in mind
\begin{equation}\label{trigav}
  \vid{\sin^2t} = \vid{\cos^2t} = 1/2, \quad \vid{\sin t\cos t} = 0,
\end{equation}
we obtain the following expression of the average flow:
\begin{equation}\label{avflow}
  \vid{J} = (k/2)\left[1 + A\cos(kx)\right].
\end{equation}
We see that by adding to the running wave homogeneous oscillations of some amplitude $A$,
we obtain a modulation of the flow of energy by the same amplitude $A$. When the amplitude
exceeds one ($A>1$), then the local backflow appears.

\section{Green's function}

The possibility of a backflow in the case of two cosine waves, discussed in the
previous section, suggests the study of more complex superpositions, including packages.
That is why we will consider the limiting case of short excitation, namely, the reaction of
the string to the instant point impact, which is described by the following equation:
\begin{equation}\label{GrynoLg}
  G_{tt} - G_{xx} + G = \delta(x)\delta(t),
\end{equation}
and boundary conditions
\begin{subequations}\label{ksps}
\begin{eqnarray}
  G(\pm\infty, t) &=& 0, \\
  G(x,t)|_{t<0} &=& 0, \\
  G_t(x,t)|_{t<0} &=& 0.
\end{eqnarray}
\end{subequations}
Actually it is the equation for Green's function of our basic Eq.~(\ref{bdlg1}).

\begin{figure}[!ht]
\includegraphics[width=60mm]{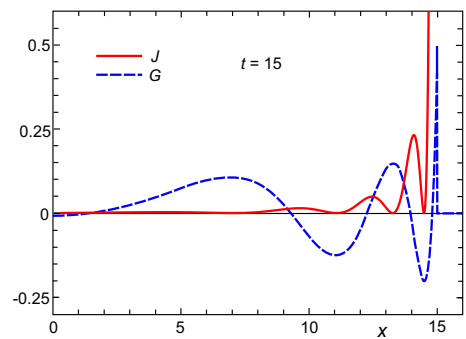}
\caption{Green's function (dashed blue curve) and the corresponding energy flow (solid red curve).}
\label{fig4}
\end{figure}

A successful solution of this equation can be obtained by analytical means.
Indeed, first we will replace this equation with homogeneous one, replacing
the Dirac functions by the proper initial condition:
\begin{subequations}\label{uzps}
\begin{eqnarray}
\label{uzps1}
&&  G_{tt} - G_{xx} + G = 0, \\
\label{uzps2}
&& G(\pm\infty, t) = 0, \\
\label{uzps3}
&& G(x,0) = 0, \\
\label{uzps4}
&& G_t(x,0) = \delta(x).
\end{eqnarray}
\end{subequations}
Next, we present the solution in the form of Fourier transform:
\begin{equation}\label{furje}
  G(x,t) = \frac{1}{2\pi}\int_{-\infty}^{\infty}dk\e^{\i kx}g(k,t).
\end{equation}
Now by substituting this expression into Eqs.~(\ref{uzps}), and using the Fourier
transform of Dirac function
\begin{equation}\label{deltafurje}
  \delta(x) = \frac{1}{2\pi}\int_{-\infty}^{\infty}dk\e^{\i kx},
\end{equation}
we convert the Green's function equations into the following ordinary
differential equation:
\begin{subequations}\label{papdif}
\begin{eqnarray}
\label{papdif1}
  \frac{d^2}{dt^2}g(k,t) + \left(1 + k^2\right)g(k,t) &=& 0, \\
\label{papdif2}
  g(k,0) &=& 0, \\
\label{papdif3}
  \frac{d}{dt}g(k,t)\Big|_{t=0} &=& 1.
\end{eqnarray}
\end{subequations}
It is easy to see that the solution of Eq.~(\ref{papdif1}), that satisfies the initial conditions
(\ref{papdif2}c), reads
\begin{equation}\label{gsprend}
  g(k,t) = \frac{\sin\left(t\sqrt{1+k^2}\right)}{\sqrt{1+k^2}}.
\end{equation}
Substituting it into Eq.~(\ref{furje}) we obtain the Green's function
\begin{equation}\label{gf}
  G(x,t) = \frac{1}{2\pi}\int_{-\infty}^{\infty}dk\e^{\i kx}
  \frac{\sin\left(t\sqrt{1+k^2}\right)}{\sqrt{1+k^2}}.
\end{equation}
Using the known representation of the exponential function
\begin{equation}\label{ekspo1}
  \e^{\i kx} = \cos(kx) + \i\sin(kx)
\end{equation}
and bearing in mind that only the even part of the function contributes to
integral (\ref{gf}), it is possible to further simplify this expression by
converting the integrand into a real function:
\begin{equation}\label{gff}
  G(x,t) = \frac{1}{\pi}\int_0^{\infty}dk\cos(kx)
  \frac{\sin\left(t\sqrt{1+k^2}\right)}{\sqrt{1+k^2}}.
\end{equation}
Such an integral can be found in the reference book \cite{ryzik},
and the Green's function can be presented in the following final form:
\begin{equation}\label{gff1}
  G(x,t) = \frac{1}{2}J_0(\sqrt{t^2-x^2})\Theta(t-x),
\end{equation}
where the symbol $J_0(x)$ stands for the Bessel function of the zero order,
and the symbol $\Theta(x)$ -- for the Heaviside step function.
It is shown in Fig.~\ref{fig4} by dashed blue curve.
We see that after hitting the string at the initial moment of time ($t=0$) a short pulse (precursor)
runs away with constant velocity, pulling behind itself slowly decaying waves.

\begin{figure}[!ht]
\includegraphics[width=60mm]{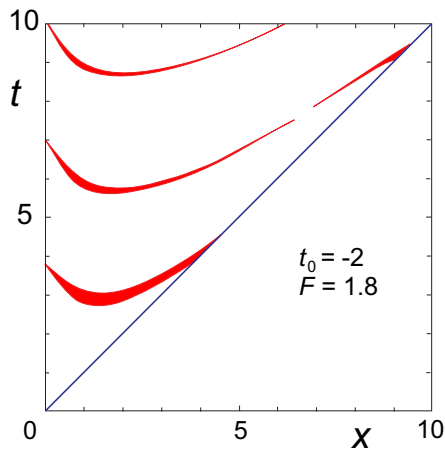}
\caption{Contour plot of energy flow in the string exited by two successive strikes,
namely, when $\delta(t)$ in Eq.~(\ref{GrynoLg}) is replaced by $\delta(t) + F\delta(t+t_0)$.}
\label{fig5}
\end{figure}

According to Eq.~(\ref{bdlg3}) we define the energy flow corresponding to the obtained
Green's function:
\begin{equation}\label{gsr}
  J(x,t) = -G_t(x,t)G_x(x,t) = \frac{xtJ_1^2(\sqrt{t^2-x^2})}{4\left(t^2-x^2\right)}.
\end{equation}
It is shown in the same Fig.~\ref{fig4} by solid red curve.
Note that the flow of energy is similar to that one seen in the case of an individual cosine wave:
it has zeros at certain wave phases, corresponding to the extremes of the wave amplitude.
Therefore, it should be expected that in the case of two successive strikes,
the propagating excitation will possibly demonstrate a local backflow.
So in fact it is, which is demonstrated by Fig.~\ref{fig5}.
The areas of the $xt$-plane where the backflow is observed are shown in red.

Comparing the results obtained in the last two sections, it may be concluded
that a local backflow of energy is possible when various excitations interfere.
The backflow manifests itself in the greatest way when two cosine waves interfere
with each other.

\section{Harmonic oscillator}

The wave systems considered in previous sections were infinite. Trying to detect the backflow
experimentally, however, one has to deal with the finite system, in which the reflection of
the wave from its edges either suppresses the flow of energy, or distorts it significantly.
Therefore, considering the finite systems one has to modify the problem, including both
the wave generation at one edge of the chain and the wave absorption on the other one,
for instance, including the dissipating force (say, friction)
into equation for the last mass in the chain shown in Fig.~\ref{fig1}.

\begin{figure}[!ht]
\includegraphics[width=40mm]{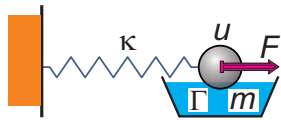}
\caption{Harmonic oscillator.}
\label{fig6}
\end{figure}

Such modification does not complicate the consideration significantly and actually leaves
a suitable result concerning the backflow, if the dissipation is small enough, almost
insensitive to the length of 1D lattice. In support of the words said in this section,
we will look at an ultra-short chain consisting only of one mass, which is both the
beginning and the end of this chain. Therefore, this mass is simultaneously under the
influence of external force and dissipating one as well.
This simple system is shown in Fig.~\ref{fig6}.
In fact, it is a harmonic oscillator, the coordinate of which satisfies the following equation:
\begin{equation}\label{VienJud}
  m\ddot{u} + \Gamma\dot{u} + \kappa u = F(t).
\end{equation}
Here the symbol $F(t)$ stands for the external force, and the dissipation is is characterized
by the friction coefficient $\Gamma$.
In presenting this well-known problem, we aim to illustrate the assertion stated in the
Introduction that the backflow of energy can appear in any oscillating system if the
energy flow can be split in several channels.

For the sake of convenience let's again introduce the dimensionless variables
\begin{equation}\label{mast1}
  t \to \Omega^{-1}t, \; \Omega = \sqrt{\kappa/m}, \; \gamma = \Gamma/m\Omega,
  \; u \to uF/m\Omega^2,
\end{equation}
(here, the symbol $F$ stands for the characteristic amplitude of external force,
namely $F(t) = Ff(t)$) and rewrite Eq.~(\ref{VienJud}) as follows:
\begin{equation}\label{baz}
  \ddot{u} + \gamma\dot{u} + u = f(t).
\end{equation}
Now following the ideas of Sec.~\ref{sec_ensr}, we multiply Eq.~(\ref{baz})
by $\dot{u}$ and get the standard energy conservation law:
\begin{equation}\label{entv}
  \frac{d}{dt}\m{E} = \frac{d}{dt}\left(\frac{1}{2}\dot{x}^2 + \frac{1}{2}x^2\right)
  = \dot{x}f - \gamma\dot{x}^2,
\end{equation}
which shows that the total energy of the oscillator $\m{E}$ changes due to the
incoming flow of energy
\begin{equation}\label{forcein}
  J = \dot{u}f
\end{equation}
from the external force and the leaking flow
\begin{equation}\label{dissout}
  D = \gamma\dot{u}^2,
\end{equation}
flowing to the thermostat represented by the water vessel in Fig.~\ref{fig6}.
Both of these quantities indicate the main flow of energy through mass
from pumping force to thermostat.

Recall that our goal is to consider the system with possible branching of the flow.
Exactly this type of branching appears in the oscillator in question.
Indeed, if we write down separately the balance for the kinetic energy of the mass
and the potential energy of the spring
\begin{subequations}\label{rutspyres}
\begin{eqnarray}
  \frac{d}{dt}T &=& \frac{d}{dt}\left(\frac{1}{2}\dot{u}^2\right) = \dot{u}\ddot{u}
  = \dot{u}f - u\dot{u} - \gamma\dot{u}^2, \\
  \frac{d}{dt}V &=& \frac{d}{dt}\left(\frac{1}{2}u^2\right) = u\dot{u},
\end{eqnarray}
\end{subequations}
we see that in addition to the above mentioned direct flow of energy,
the energy flow
\begin{equation}\label{Wdef}
  W = u\dot{u}
\end{equation}
appears, which takes off the energy from the main flow to the spring.
Let's check if there is a possibility of the negative flow $J<0$ of energy,
which could be called an analog to the previously considered backflow
in the infinite chain.

First, we consider the case when the force is of a single cosine type component:
\begin{equation}\label{vienaskos}
  f(t) = \cos(\omega t) = \Re \e^{-\i\omega t}.
\end{equation}
The standard stationary solution of Eq.~(\ref{baz}) reads
\begin{subequations}\label{xvvien}
\begin{eqnarray}
  u &=& \Re \frac{\e^{-\i\omega t}}{1-\omega^2 -\i\gamma\omega} \nonumber \\
  &=& \frac{(1 - \omega^2)\cos(\omega t) + \gamma\omega\sin(\omega t)}
  {(\omega^2 - 1)^2 + \gamma^2\omega^2}, \\
\label{xvvien2}
  \dot{u} &=& a(\omega)\cos(\omega t) + b(\omega)\sin(\omega t),
\end{eqnarray}
\end{subequations}
where the amplitudes
\begin{subequations}\label{abcoefdef}
\begin{eqnarray}
  a(\omega) &=& \frac{\gamma\omega^2}{(\omega^2 - 1)^2 + \gamma^2\omega^2}, \\
  b(\omega) &=& \frac{\omega(\omega^2-1)}{(\omega^2 - 1)^2 + \gamma^2\omega^2}
\end{eqnarray}
\end{subequations}
actually are the real and imaginary parts of Lorenzian.

\begin{figure}[!ht]
\includegraphics[width=60mm]{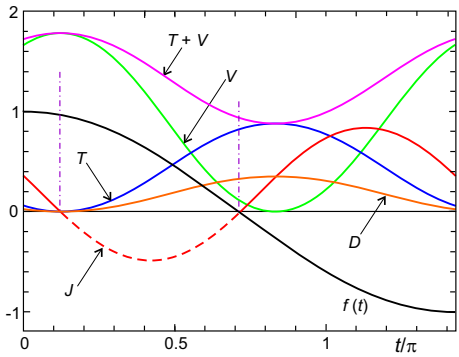}
\caption{Energies and flows given by formulas (\ref{entv}), (\ref{rutspyres}),
(\ref{vienaskos}) and (\ref{viensrautai1}).}
\label{fig7}
\end{figure}

Using the obtained expressions for the coordinate of the oscillator and its velocity,
we determine the energy flows that are of interest to us:
\begin{subequations}\label{viensrautai}
\begin{eqnarray}
\label{viensrautai1}
 J &=& \frac{a(\omega)}{2}
  + \frac{\omega\sin(2\omega t + \ph_J)}{2\sqrt{(\omega^2-1)^2 + \gamma^2\omega^2}}, \\
\label{viensrautai2}
 D &=& \frac{a(\omega)}{2}\left[1 + \sin(2\omega t + \ph_D)\right], \\
\label{viensrautai3}
 W &=& \frac{\omega\cos(2\omega t + \ph_D)}{2\left[(\omega^2-1)^2 + \gamma^2\omega^2\right]},
\end{eqnarray}
\end{subequations}
where
\begin{subequations}\label{phidef}
\begin{eqnarray}
  \ph_J &=&  \arctan\left[a(\omega)/b(\omega)\right], \\
  \ph_D &=& \arctan\left[\frac{\gamma^2\omega^2 - (\omega^2-1)^2}{2\gamma\omega(\omega^2-1)}\right].
\end{eqnarray}
\end{subequations}
These flows together with kinetic $T$ and potential $V$ energies are shown in Fig.~\ref{fig7}.

Paying attention to the obtained formulas and Fig.~\ref{fig7}, we can note some interesting points.
The averaged flows over oscillations with double frequency
\begin{equation}\label{vidut}
  \vid{J} = \vid{D} = a(\omega)/2
\end{equation}
are positive and indicate that the main flow of energy goes from the external force to the thermostat.
The magnitude of this flow coincides with the amplitude of the oscillating part of the
dissipative flow $D$ (see the second term in the right hand side of Eq.~(\ref{viensrautai2})).
This confirms that this flow is always positive, which it should be, because the dissipative
part of the system has no source of energy.
Meanwhile, the pumping energy (flow $J$) can acquire negative values, because the amplitude
of the second term in the right part of Eq.~(\ref{viensrautai1}) is always larger than
its first term, except for the point of resonance ($\omega=\pm 1$), where both terms become equal.
This negative part of the pumping energy is highlighted by the dashed part of the $J$ curve.
This is a rather trivial property of the forced oscillator, which is often overlooked,
because most often only the average power absorption is of interest.
In our case, as mentioned above (see Eq.~(\ref{vidut})), it is always positive.

To get a negative average pumping power that manifests itself over a long period of time,
one needs to try to excite the oscillator with a force consisting of two periodic functions
of close frequency, for instance, postulating the external force as follows:
\begin{equation}\label{dvijeg}
\begin{split}
  f(t) &= \cos\left[\right(\omega - \Delta\omega/2)t] + \cos\left[\right(\omega + \Delta\omega/2)t] \\
  &\equiv f_- + f_+, \qquad \Delta\omega \ll \omega.
\end{split}
\end{equation}
Let us note that such a double excitation of the oscillator is a good analog to the previously
considered problem of two cosine wave propagation in an infinite chain.
Here and in further terms for brevity we will use $\pm$ indexing of functions instead of
indicating explicitly their dependence on shifted frequencies $\omega_{\pm}=\omega\pm\Delta\omega/2$.
Therefore, according to Eq.~(\ref{forcein}), the energy pumped into the
oscillator by the exciting force can be represented in the following form:
\begin{equation}\label{pump}
  J = \dot{u}_-f_- + \dot{u}_+f_+ + \dot{u}_-f_+ + \dot{u}_+f_-,
\end{equation}
where the expressions for velocities $\dot{u}_{\pm}$ follow from Eq.~(\ref{xvvien2})
by proper replacement of frequency.

To calculate the average of this equation, one needs the following averages
of trigonometry functions products:
\begin{equation}\label{apytlg1}
\begin{split}
&  \vid{\cos^2\left[(\omega\pm\Delta\omega/2)t\right]} = 1/2, \\
&  \vid{\cos\left[(\omega\pm\Delta\omega/2)t\right]\sin\left[(\omega\pm\Delta\omega/2)t\right]} = 0, \\
&  \vid{\cos\left[(\omega\pm\Delta\omega/2)t\right]\cos\left[(\omega\mp\Delta\omega/2)t\right]} \\
&  = \vid{\left[\cos(\omega t)\cos(\Delta\omega t/2) \mp \sin(\omega t)\sin(\Delta\omega t/2)\right] \\
&  \times\left[\cos(\omega t)\cos(\Delta\omega t/2) \pm \sin(\omega t)\sin(\Delta\omega t/2)\right]} \\
&  = \left[\cos^2(\Delta\omega t/2) - \sin^2(\Delta\omega t/2)\right]/2
   = \frac{1}{2}\cos(\Delta\omega), \\
&  \vid{\cos\left[(\omega\pm\Delta\omega/2)t\right]\sin\left[(\omega\mp\Delta\omega/2)t\right]}
   = \mp\frac{1}{2}\sin(\Delta\omega), \\
\end{split}
\end{equation}
which can be calculated assuming that the averaging over fast oscillations with frequency $\omega$
does not affect the slow beats.

The averaging of two first terms in eq.~(\ref{pump}) is simple: it leads to
\begin{equation}\label{dupirmJ}
\begin{split}
&  \vid{\dot{u}_-f_-} + \vid{\dot{u}_+f_+} = \left(a_- + a_+\right)/2 \\
&  \equiv \left[a(\omega - \Delta\omega/2) + a(\omega + \Delta\omega/2)\right]/2 = A(\omega),
\end{split}
\end{equation}
that gives only a permanent background.
Meanwhile, the averaged two remaining terms demonstrate the beatings:
\begin{subequations}\label{likdu}
\begin{eqnarray}
  \vid{\dot{u}_-f_+} &=& \left[a_-\cos(\Delta\omega t) - b_-\sin(\Delta\omega t)\right]/2, \\
  \vid{\dot{u}_+f_-} &=& \left[a_+\cos(\Delta\omega t) + b_+\sin(\Delta\omega t)\right]/2.
\end{eqnarray}
\end{subequations}
Putting all four of the above terms together, we get the final expression of the pumping power:
\begin{equation}\label{vidfin}
  \vid{J} = A(\omega)\left[1 + \cos(\Delta\omega t)\right] + B(\omega)\sin(\Delta\omega t),
\end{equation}
where
\begin{equation}\label{Bdef}
  B(\omega) = \left(b_+ - b_-\right)/2.
\end{equation}

\begin{figure}[!ht]
\includegraphics[width=60mm]{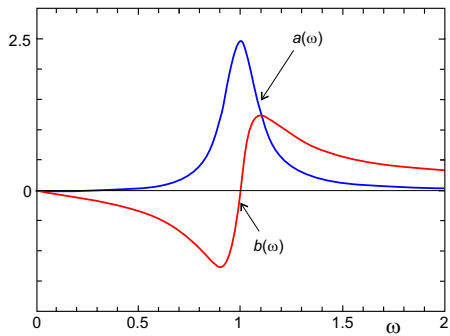}
\caption{Auxiliary functions (\ref{abcoefdef}).}
\label{fig8}
\end{figure}

\begin{figure}[!ht]
\includegraphics[width=60mm]{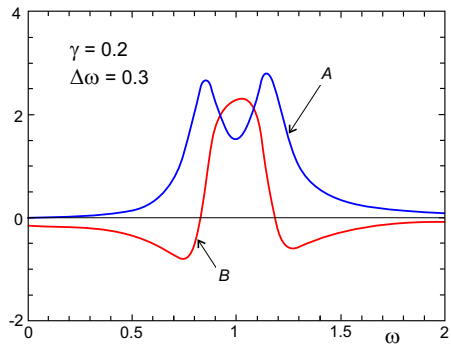}
\caption{Functions $A$ and $B$.}
\label{fig9}
\end{figure}

It is remarkable that averaging Eq.~(\ref{vidfin}) over the period of beats
($2\pi/\Delta\omega$) we obtain $\vid{J}=A$, what coincides with the average dissipation flow.
Another good fact for us is that the amplitude of the oscillating part of Eq.~(\ref{vidfin})
is equal to $\sqrt{A^2+B^2}$ and is always larger than its constant part $A$.
Thus, there is always a certain phase interval where the mean pumping flow is negative,
i.~e.~there is an analog to backflow which we are looking for.
This backflow is larger and appears in the wider phase interval, the greater is the
the difference $(B-A)$. To get an idea when this is possible, let's take a
look at Fig.~\ref{fig8},
where the frequency  dependencies of auxiliary functions (\ref{abcoefdef}) are shown.
Because function $B$ consists of a difference of two functions $b$ at close frequencies
(it can be seen as a derivative of function B by frequency), its value is the largest in the
resonance region ($\omega\approx 1$), as it is seen in Fig.~\ref{fig8}.
We can see this in Fig.~\ref{fig9} where the functions $A$ and $B$ are shown in the
case of a concrete detuning $\Delta\omega=0.3$.
We see that the function $B$ has the largest value (exceeding the value of function $A$)
at the resonance point ($\omega=1$).

\begin{figure}[!ht]
\includegraphics[width=60mm]{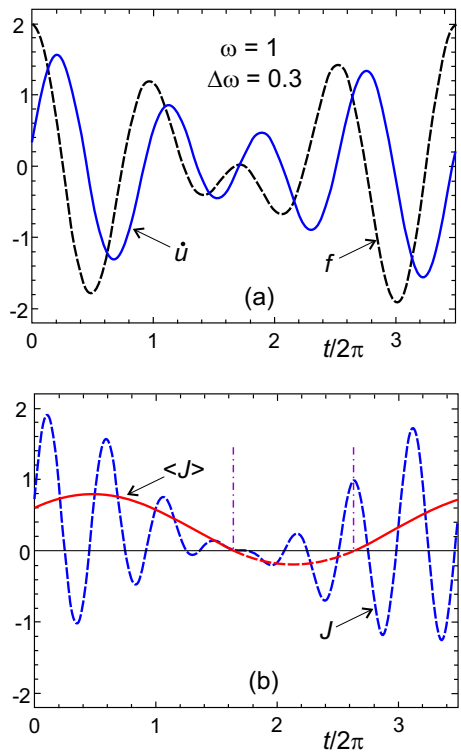}
\caption{Oscillator excited by external force consisting of two periodic components:
(a) -- temporal behavior of force and velocity, (b) -- pumping energy and its averaged value.}
\label{fig10}
\end{figure}

Fig.~\ref{fig10} sums up this simple reasoning.
Here the temporal dependencies of the above mentioned functions are shown.
In part (a) of this graph, the external force (\ref{dvijeg}) is shown by a black dashed
curve, and the blue solid curve depicts the velocity $\dot{u}$ of the oscillator,
calculated according to  Eq.~(\ref{xvvien2}).
Both of these curves clearly demonstrate the beats, what should be the case due to
the proximity of the frequencies of the two cosine type components of the external force,
and the lag in the velocity phase.

In part (b) of the figure, we see the corresponding energy flows caused by the driving force:
the blue dashed curve shows the instantaneous value of the flow $J$, and the solid red -- its
averaged value according to Eq.~(\ref{vidfin}).

We see that the instantaneous flow $J$ has negative values at some interval of time of each
period of the mean frequency, as was the case with single-component external force.
The averaged flow also demonstrates negative values (see the dotted part of the red curve),
that take up a significant portion of the beating period.

\section{Conclusions}

We have shown that in the classical system described by differential equations with a spectrum
that has a cut-off frequency, there is the possibility of a backflow of energy localized both
by coordinate and time as in the known case of quantum mechanics in literature.
The main condition for the appearance of this backflow is the wave dispersion, which appears
due to the presence of local mods in the medium of propagation.
The most favorable opportunity occurs when two waves of slightly different frequencies
are propagating. In this case, the backflow occurs both in the microscopic description
and using averages.

We illustrated these claims by considering the simplest mechanical system consisting of a
chain of masses connected by springs between the nearest neighbours and by additional
springs attached to stationary supports,
which actually ensures the appearance of the local mods.

Concerning propagation of packages, we studied Green's function for the above system and
showed that the conditions for backflow are not so favorable. However, this is also
possible in the case with two consecutive well-coordinated strikes.

We claim that the backflow appears as a result of the branching of energy flows, and
demonstrate this on the example of the simplest model of forced damped oscillator.
Here the backflow (the return of energy to external power) is quite impressive.
The instantaneous backflow is already observed at the excitation by a single cosine type
force component. In the case of two cosine type force components of close frequency
the averaged backflow takes place.

The proposed mechanical chain can be carried out in an experimentally simple way.
The characteristic frequencies in the Hz range are easy to observe in a complete analogy
with the S\"{u}sstrunk and Huber experiments \cite{sus15}, where the pendulum system was used to
demonstrate edge mods in quantum mechanics.

\begin{figure}[!ht]
\includegraphics[width=35mm]{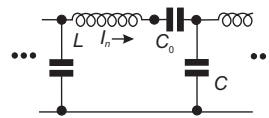}
\caption{Cell of transmission line.}
\label{fig11}
\end{figure}

It is also worth noting that from a mathematical point of view the mechanical system
considered is equivalent to an electric one, namely, a transmission line made up of cells,
shown in Fig.~\ref{fig11}.
Using Kirchhoff's equations, it is easy to show that the currents $I_n$ of  inductancies
$L$ are described by the same type Eq.~(\ref{ChJud2}) as the coordinates of
the masses of the considered chain. We believe that using such a transmission line it is possible
to observe the considered phenomena of backflow in MHz range.

%

\end{document}